\documentclass[%
 aip,
 amsmath,amssymb,
 reprint,%
]{revtex4-1}

\usepackage{graphicx}
\usepackage{dcolumn}
\usepackage{bm}

\usepackage[utf8]{inputenc}
\usepackage[T1]{fontenc}
\usepackage{mathptmx}
\usepackage{etoolbox}

\makeatletter
\def\@email#1#2{%
 \endgroup
 \patchcmd{\titleblock@produce}
  {\frontmatter@RRAPformat}
  {\frontmatter@RRAPformat{\produce@RRAP{*#1\href{mailto:#2}{#2}}}\frontmatter@RRAPformat}
  {}{}
}%
\makeatother
\begin{document}

\preprint{AIP/123-QED}

\title{Lagrangian mixing of pulsatile flows in constricted tubes}
\author{N. Barrere}
\email{nbarrere@fisica.edu.uy}
\affiliation{PDU Ciencias Físicas, Centro Universitario Regional Este, Universidad de la República, Uruguay}
\affiliation{Instituto de Física, Facultad de Ciencias, Universidad de la República, Uruguay}

\author{J. Brum}%
 \affiliation{Instituto de Física, Facultad de Ciencias, Universidad de la República, Uruguay}

\author{M. Anzibar}
 \affiliation{Instituto de Física, Facultad de Ciencias, Universidad de la República, Uruguay}

\author{F. Rinderknecht}
\affiliation{Instituto de Física, Facultad de Ciencias, Universidad de la República, Uruguay}

\author{G. Sarasúa}
\affiliation{Instituto de Física, Facultad de Ciencias, Universidad de la República, Uruguay}

\author{C. Cabeza}
\affiliation{Instituto de Física, Facultad de Ciencias, Universidad de la República, Uruguay}
\date{\today}

\begin{abstract}
In this work several lagrangian methods were used to analyze  the mixing processes in an experimental model of a constricted artery under a pulsatile flow. Upstream Reynolds number $Re$ was changed between 1187 and 1999, while the pulsatile period $T$ was kept fixed at 0.96s. Velocity fields were acquired using Digital Particle Image Velocimetry (DPIV) for a region of interest (ROI) located downstream of the constriction. The flow is composed of a central jet and a recirculation region near the wall where vortex forms and sheds. To study the mixing processes, finite time Lyapunov exponents (FTLE) fields and concentration maps were computed. Two lagrangian coherent structures (LCS) responsible for mixing and transporting fluid were found from FTLE ridges. A first LCS delimits the trailing edge of the  vortex, separating the flow that enters the ROI between successive periods. A second LCS delimits the leading edge of the vortex. This LCS concentrates the highest particle agglomeration, as verified by the concentration maps. Moreover, from the particle residence time maps (RT) the probability for a fluid particle of leaving the ROI before one cycle was measured. As $Re$ increases, the probability of leaving the ROI increases from 0.6 to 0.95. Final position maps $r{_f}$ were introduced to evaluate the flow mixing between different subregions of the ROI. These maps allowed us to compute an exchange index between subregions, $\overline{\mathrm{EI}}$, which shows the main region responsible for the mixing increase with $Re$.
	
	Finally by integrating the results of the different lagrangian methods (FTLE, Concentration maps, RT and $r_f$ maps), a comprehensive description of the mixing and transport of the flow was provided. 

\end{abstract}

\maketitle

\section{Introduction}\label{sec1}

Hemodynamics is a very important research topic for its implications in cardiovascular diseases  \cite{Vlachopoulos2011, naghavi}. One of the most dangerous diseases in arteries is atherosclerosis \cite{atlas_cardiovascular}, characterized by the aggregation of fat and cholesterol which leads to a partial obstruction of the arterial lumen. An accurate description of blood flow under these conditions has become key to understand the initiation and progression of such diseases. In this context, mixing among regions within the artery is known to play a major role because of its dependence to disturbances in blood flow \cite{Liu2011, shadden_applied_lagrangian}. 

The mechanisms that govern the mixing processes are complex and diverse, including advection, diffusion, and turbulence. Since the early works of Winter and Nerem \cite{Winter1984}, several authors \cite{peacock, trip,Brindise2018, Xu2020} have studied the transition to turbulence. For instance, the works of Jain \cite{Jain2019, Jain2022} serve as a benchmark in studying the transition to turbulence in an oscillatory flow through a rigid pipe with stenosis. These works showed, among other things, that the pulsatile frequency variation does not modify the critical Reynolds for transition. However, mixing occurs in a non-turbulent regime, and a comprehensive description of it is still missing.

Lagrangian methods are based on particle tracking and they provide more insight into the flow topology than other widely used Eulerian fields. Therefore, they offer a framework to understand mixing processes and are essential in quantifying mixing and other features of pulsatile flows \cite{shadden_applied_lagrangian}. Particularly,  finite-time Lyapunov exponents (FTLE) \cite{haller_2000,haller_2000_chaos,shadden-tesis,shadden_2005} are used to identify Lagrangian coherent structures (LCS) that act as material barriers organizing the flow. 

FTLE have been successfully applied in hemodynamic models to study steady and pulsatile flows \cite{arzani_2012, shadden_applied_lagrangian,badass_2017}. Vétel et al. \cite{vetel} studied the flow near the bifurcation of a carotid artery, 
identifying the formation of structures, such as vortices, through FTLE fields. 
Espa et al. \cite{espa} have applied these techniques to study the flow within the left ventricle. Similarly, Arzani and Shadden \cite{arzani_2012} calculated FTLE fields from numerical data to study the transport of vortices, determining the areas in which vortices are detached, and characterizing the mixing in aortic aneurysms. However, the description of how mixing occurs could be complemented by other tools such as concentration maps. The combination of these lagrangian techniques would enable the measurement of  fluid from each structure taking part in the mixing.  

Another widely-used Lagrangian method is the distribution of residence time (RT) and RT  maps \cite{residence_time_camassa_91,goulliart_2011, monsen_2002, Rayz_2010}, which represent the time a given particle remains in a specific region. Several authors have used RT to study blood-flow mixing in the left ventricle \cite{Seo2013, Long2013,dilabbio_2018,badass_2017}. Di Labbio et al. \cite{dilabbio_2018} studied an experimental model of a left ventricle with regurgitation. The authors observed that this condition increases the RT of the flow inside the ventricle, with the potential risk of suffering from arrhythmia or other cardiovascular diseases. Other studies have reported that RT maps can be used to analyze how long platelets remains in a specific region \cite{Rayz_2010,fuster_2005}.

Particle residence time (PRT) is a path-dependent metric which quantifies the time a fluid parcel spends in a region of interest \cite{jeronimo2019,jeronimo2020}. However, PRT differs from RT since it tracks fluid tracers experimentally while RT is based on advecting particles from the eulerian velocity field. The work of Jeronimo et al. \cite{jeronimo2019} studied PRT of a flow in a model of stenosed artery. This allowed the identification of  particles initial region  and subsequent trajectory, categorizing the flow in reverse flow, jet, recirculating flow and others. However, a map showing the exact final positions of particles would allow a more accurate description of the mixing. 

In the present work, Lagrangian methods were used to study the mixing of a pulsatile flow in an experimental model of an artery with stenosis. The study focused on the region downstream of the constriction, analyzing the mixing processes by means of FTLE fields and concentration maps. The FTLE fields allowed the identification of the LCS that delimits two regions with well-differentiated dynamics: the vortices against the wall and the central jet. Overall, we explained the characteristics of the flow and the mechanism of mixing using FTLE fields and Concentration maps. Quantitative analysis of mixing is not easy since it usually depends on each specific application; however, in this work, we present Lagrangian descriptors capable of measuring mixing. The final state of the mixing was analyzed through RT maps and $r_f$ maps, which show the final position of the particles. Two relevant parameters were computed from these maps: the probability of flow leaving the ROI and the $\overline{\mathrm{EI}}$ index, which measures flow exchange between subregions within the ROI. Finally, a detailed discussion is given on the results obtained from the FTLE fields, $C$ maps, RT, and $r_f$ and how they complement each other.


The work is organized as follows. In sec. \ref{sec:M&M} we describe the experimental setup and the computation of aforementioned lagrangian methods. In sec. \ref{sec:res} we briefly show the results obtained by each lagrangian method and in sec. \ref{sec:discus} we discuss the results altogether. Finally, in sec. \ref{sec:conc} we present the conclusions. 

\section{Materials and methods} \label{sec:M&M}
\subsection{Experimental setup}
\begin{figure*}[htb]
	\resizebox{1\textwidth}{!}{%
		\includegraphics{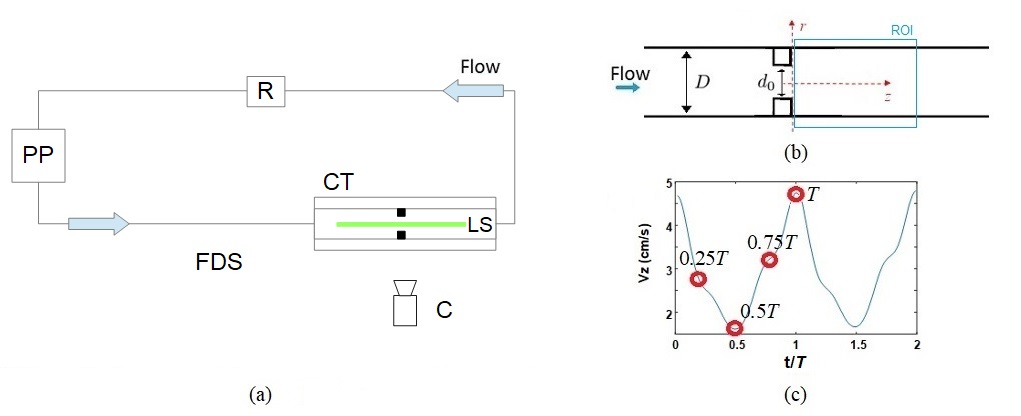}
	}
	\centering
	\caption{	
		(a) Experimental setup. \textbf{FDS:} Flow development section \textbf{PP:} Programmable pump; \textbf{R:} Reservoir; \textbf{LS:} Laser sheet; \textbf{CT:} constricted tube; \textbf{C:} camera. (b) Cross-sectional view of a tube of diameter $D = 2.6 \pm 0.1$ cm and an annular constriction of diameter $d_0=1.6\pm 0.1$cm; (c) $V_z$ at the inlet and $r = 0$ for $Re$ =1187. Red circles indicates selected time locations.}
	\label{fig:vz_centerline}       
\end{figure*}

The experimental set up is shown in fig. \ref{fig:vz_centerline}. Pulsatile flow generated by a programmable pump (PP) was led into a constricted tube (CT). This constricted tube consisted of a transparent acrylic tube of diameter $D=2.6 \pm 0.1$ cm containing a hollow, cylindrical annular constriction with an internal diameter $d_0=1.6 \pm 0.1$ cm, which reduced the area by a factor of 39\% \cite{Barrere2020}. 

A flow development section (FDS) was connected upstream CT to ensure developed flow \cite{Barrere2020}. The region of interest, ROI, was defined as -0.5$<r/D<$0.5, 0$< z/D <$1.5, see fig\ref{fig:vz_centerline}(b). Digital Particle Image Velocimetry (DPIV) technique was used to acquire the velocity fields  \cite{piv1}. Distilled water was seeded with neutrally buoyant 50$\mu$m  particles. A 1 W Nd:YAG laser was used to illuminate a 2 mm-thick section of the tube. Sixteen pulsatile cycles were acquired at a frame rate of 180Hz using a CMOS camera (Pixelink, PL-B776F). The velocity field was finally computed using PIVlab software \cite{Thielicke2019} with $32 \times 32$ pixel$^2$ windows and an overlap of 8 pixels in both directions. 

Upstream Reynolds number was defined as $Re=Dv_u/\nu$, where $v_u$ is the upstream peak velocity at the centerline ($r=0$) and $\nu$ is the kinematic viscosity of water ($\nu \mathrm{=1.0\times10^{-6} m^2/s}$ ). Four experiments were performed with Reynolds values $Re$=[1187, 1427, 1625, 1999] and $T$=0.96s for all cases. Axial velocity at $z = 0$ and $r = 0$ as function of time (fig. \ref{fig:vz_centerline}(c)) was used as time reference for all experiments. The decelerating phase was defined as $0<t<0.5T$, whereas the accelerating phase was defined as $0.5T<t<T$.

\subsection{Lagrangian Coherent Structures detection}\label{sec:ftle}
LCS are defined as manifolds that act as flow separatrices, i.e. material barriers \cite{haller_2000, haller_2000_chaos}, and they correspond to the ridges of the FTLE fields \cite{shadden_2005, shadden-exp}. The FTLE fields measure the exponential separation between initially adjacent particles during a finite time interval ($t_0, t_0+\tau$) \cite{haller_2000, haller_2002}. This means that, if $\left \| \delta \mathbf{x}_0 \right \|$ is the initial separation between two fluid particles, the maximum separation will be given by

\begin{equation}
\left \| \delta \mathbf{x} \right \|_{\textrm{max}}=e^{\left \| \tau \right \|\Lambda}\left \| \delta \mathbf{x}_0 \right \|
\label{ftle}
\end{equation}

where $\Lambda$ is the FTLE field calculated in the $(t_0, t_0+\tau)$ time interval. To compute FTLE fields, particles trajectories $\mathbf{x}(t)$ were computed by numerically solving $\mathbf{\dot{x}}\left ( t \right )=\mathbf{v}\left( \mathbf{x}, t \right )$.

To this end, the numerical procedure sub-divides the grid of the two-dimensional velocity field and seeds $3 \times 3$ artificial particles in each of its cells. For simplicity, artificial particles will be named particles throughout this work. 
Particles were then advected following a 4th order Runge-Kutta method and a cubic interpolation. Further refinement of the grid produced negligible changes in the results. The FTLE fields are then calculated based on the Cauchy-Green tensor ($CG$), given that $\left \| \delta \mathbf{x} \right \|_{\textrm{max}}$ is aligned with the eigenvector with maximum eigenvalue, $\lambda_{\textrm{max}}(CG)$, of the $CG$ tensor. This means that $\left \| \delta \mathbf{x} \right \|_{\textrm{max}}=\sqrt{\lambda_{\textrm{max}}(CG)}\left \| \delta \mathbf{x}_0 \right \|$. The FTLE fields is calculated as

\begin{equation}
\Lambda=\frac{1}{\left \| \tau \right\|} \mathrm{ln}(\sqrt{\lambda_{\textrm{max}}(CG)})
\label{ftle_practica}
\end{equation}

Trajectories can be integrated forward or backward in time. If $\tau>0$, FTLE fields measure separation forward in time whereas if $\tau<0$ FTLE fields measure separation backward in time. These FTLE fields were named $\Lambda^+$ and $\Lambda^-$ and they highlight the repelling and attracting manifolds respectively \cite{haller2015, haller_2002}.

The integration time $\tau$ was determined based on experimental data such as the time it takes for a particle located in the vortex boundary to go through half of that boundary contour, yielding a value of $\tau$= 0.25$T$. To identify the ridges associated to LCS, all FTLE values below a threshold of 50\% of the maximum FTLE value were discarded.

\subsection{Concentration, Residence time and final position maps}
\label{sec:teo_rt}

For this study, the Péclet number $Pe=v_uD/d$, which compares advection and diffusion, is of the order of $Pe>$1$\times$10$^6$, where $d$ is the mass diffusivity (10$^{-9}$m$^2$/s for distilled water). Consequently, the mixing and transport are primarily due to advection. Therefore, we calculated concentration maps $C$, residence time maps RT and final position maps of particles $r_f$.

Instantaneous concentration maps $C$ were calculated from particle trajectories computed in section \ref{sec:ftle}. To this end, we get the exact particles coordinates in each timestep and assign its location to its corresponding DPIV cell. Finally, we count the number of particles in each DPIV cell \cite{cagney_2016}. 

\subsubsection{Residence Time maps and probability of a particle leaving the ROI}

For the RT maps, particles were initially distributed uniformly throughout the ROI with the same spatial resolution used to compute the FTLE fields. Then, particles were advected from an initial time $t_0$ during a timespan of two pulsatile cycles. Each pixel of the RT map corresponds to the fluid particle initial position and the pixel value (i.e. color scale) is given by the time spent by this particle within the ROI. With this computation already used in \cite{badass_2017}, the RT map will depend on the initial time $t_0$. To make this dependence explicit, RT maps will be denoted RT($t_0$). 


We computed $\mathrm{\overline{RT}}$ by averaging over $t_0$ in the time interval $(0,T)$. The $\mathrm{\overline{RT}}$ map gives the mean residence time for any point of the ROI regardless of the initial time $t_0$. Finally, the proportion of particles which leave the ROI before a pulsatile cycle, $P\left(\mathrm{\overline{RT}}<T \right)$, is computed by counting the particles with $\mathrm{\overline{RT}}$ smaller than a pulsatile period \cite{goulliart_2011,Danckwerts1952}.

\subsubsection{Final position maps and particle exchange index between regions}

The first step to evaluate the mixing inside the ROI was to compute the final position maps ($r_f$) where each pixel corresponds to the fluid particle initial position and the color value will be given by its final position in $r$. Since particles left the ROI through the boundary at $z/D=1.5$, we considered its final positions only in $r$ coordinate. In this work we divided the whole ROI in two subregions $R_1$ and $R_2$ in order to measure the particle exchange between these subregions. 




Thus, we evaluate the particles which leave through a specific subregion. For instance, considering a time interval $(t_0, t_0+t_f)$, the total number of particles leaving the ROI through $R_1$ is $N_1^{t_f}$ and can be written as $N_1^{t_f}=N_1^{t_o \rightarrow t_f}+N_{2\rightarrow 1}^{t_o \rightarrow t_f}$. Where $N_{1}^{t_o \rightarrow t_f}$ is the number of particles  in $R_1$ at $t=t_0$  which left the ROI through $R_1$  and $N_{2\rightarrow 1}^{t_o \rightarrow t_f}$ is the number of particles in $R_2$ at $t=t_0$ which left the ROI through $R_1$. Normalization by $N_1^{t_f}$ yields $1=\mathrm{EI}_1 +\frac{N_{2\rightarrow 1}^{t_o \rightarrow t_f}}{N_1^{t_{f}}}$, where

\begin{equation}
\mathrm{EI}_1=\frac{N_1^{t_o \rightarrow t_f}}{N_1^{t_{f}}}    
\end{equation}

is defined as an index which measures the proportion of particles initially in $R_1$ to all particles leaving the ROI through $R_1$. Analogously, we get  $\mathrm{EI}_2=\frac{N_2^{t_o \rightarrow t_f}}{N_2^{t_{f}}}$ for subregion $R_2$.

If $\mathrm{EI}_1<1$, it means that $\mathrm{R}_1$ received particles from $\mathrm{R}_2$ while  $\mathrm{EI}_2<1$ means that  $R_2$ received particles from $R_1$. Finally, in order to get a mean rate of exchange for any time of the pulsatile cycle, we defined $\overline{\mathrm{EI}}_j$ as the mean of $\mathrm{EI}_j$ over all the initial conditions $t_0$ for $j=1,2$.

\section{Results} \label{sec:res}

\begin{figure*}[htb]
	\resizebox{0.5\textwidth}{!}{%
		\includegraphics{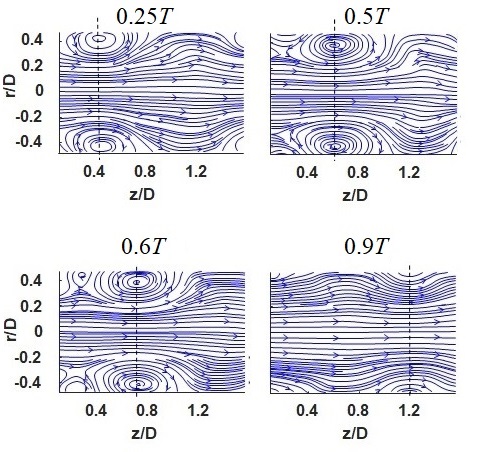}}
	\centering
	\caption{
		Streamlines for $Re$ = 1187 at $t$=[0.25 0.5 0.6 0.9]$T$. Black dashed line indicates the vortex centre.}
	\label{fig:muchas_stream1}       
\end{figure*}

Figure \ref{fig:muchas_stream1} shows streamlines for $Re=$1187 at $t=0.25T$, $t=0.5T$, $t=0.6T$ and $t=0.9T$. This flow is axisymmetric, as reported in previous studies \cite{Barrere2020, sherwin_2005,griffith,usmani}. We observed two different structures: a high-velocity jet in the central region, and a recirculation region between the jet and the walls. This recirculation region is where the vortices form and shed. During the deceleration phase, the recirculation near the constriction generates a new vortex \cite{griffith, usmani}, while the vortex generated the previous period continues travelling 
until it finally dissipates. This flow structure was consistent throughout all the experiments. A complete description of the flow structure for these experiments was previously reported in \cite{Barrere2020}. 

According to previous studies \cite{sadan, casanova, varghese_2007, trip, Xu2017}, there is no transition to turbulence for the parameters used in our work. 
In such a comparison, Reynolds and Womersley $\left(\alpha=\frac{D}{2}\sqrt{\frac{2\pi}{T\nu}}\right)$ dimensionless numbers  must be taken into consideration. For instance, the works of Cassanova \& Giddens \cite{casanova} and later confirmation made by Varghese \cite{varghese_2007} do not find transition to turbulence for an axisymmetric constriction of 75\%, peak $Re$=2240, and $\alpha$=15.6 in the immediate downstream region. Although $\alpha$=33.6 in our work, other studies \cite{Xu2017,trip,sadan} indicate that for $\alpha >12$, transition to turbulence is independent of $\alpha$.



\subsection{Lagrangian Coherent Structures} \label{sec:lcs}

\begin{figure*}[htb]
	\resizebox{0.85\textwidth}{!}{%
		\includegraphics{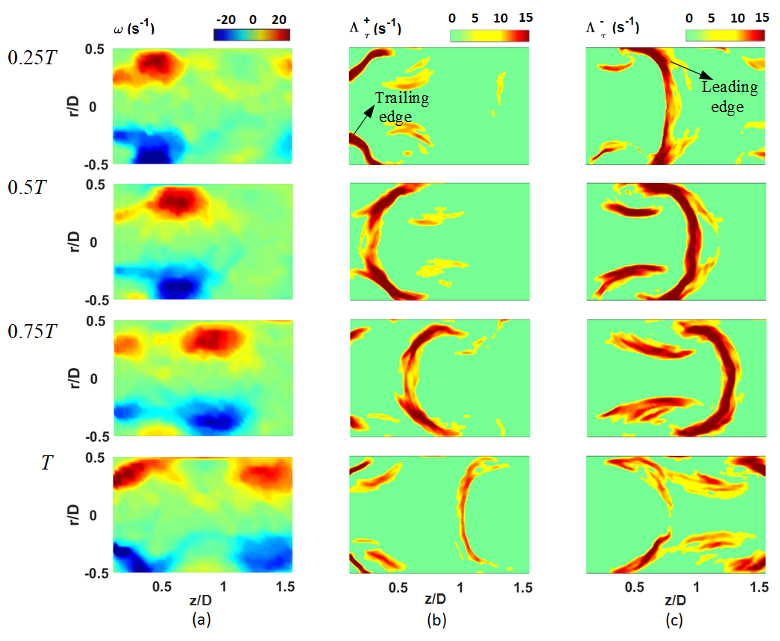}}
	\caption{For $Re$=1187, at $t$=[0.25 0.5 0.75 1]$T$ (a) vorticity field $\omega \,(s^{-1})$. Positive vorticity region(counterclockwise) against the upper wall and negative vorticity region (clockwise) against the lower wall; (b) positive FTLE fields $\Lambda^+\, (s^{-1})$;  (c) negative FTLE fields $\Lambda^-\, (s^{-1})$. Both FTLE fields were computed with $\tau$=0.25$T$. Colorbars on the head of the figure.}
	\label{fig:FTLE_barrera1}       
\end{figure*}

Figure \ref{fig:FTLE_barrera1} shows vorticity field, $\Lambda^+$ and $\Lambda^-$ fields, for $Re$=1187. Similar results were obtained for the other experiments (shown in appendix \ref{sec:app}). Figure \ref{fig:FTLE_barrera1} (a) shows four snapshots of the vorticity field $\omega(s^{-1})$ for 0.25$T$, 0.5$T$, 0.75$T$ and $T$. The vortex is already formed at $t=$0.25$T$ and its center is located approximately at $z$=0.5$D$. At $t$=0.75$T$  a second vortex starts to form near the constriction, while at $t=T$ the first vortex is close to the edge of the ROI.

Figures \ref{fig:FTLE_barrera1} (b) and (c) show the ridges of $\Lambda^+$ field and $\Lambda^-$ field, which represent the repelling and attracting LCS, respectively. In fig. \ref{fig:FTLE_barrera1} (b) a ridge of the $\Lambda^+$ field delimits the trailing edge of the vortex whereas in  fig. \ref{fig:FTLE_barrera1} (c) a ridge of the $\Lambda^-$ field delimits the leading edge of the vortex. This set of ridges move together with the vortex. Moreover, figs. \ref{fig:FTLE_barrera1} (b) and (c) show ridges parallel to the wall that separates the vortex from the central jet. Analogue LCS representing the leading edge and the trailing edge of the vortex were found for the other experiments (see appendix \ref{sec:app}). 

\subsection{Concentration maps}

\begin{figure*}[htb]
	\centering
	\resizebox{0.78\textwidth}{!}{%
		\includegraphics{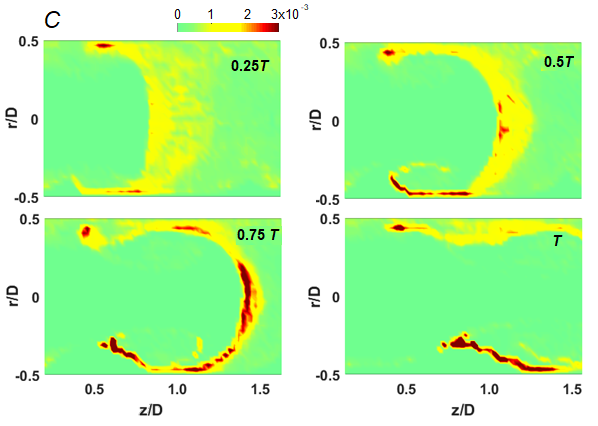}}
	\centering
	\caption{Instantaneous normalized particle concentration maps $C$ for $t=$0.25$T$, 0.5$T$, 0.75$T$ and $T$ for $Re$=1187.}
	\label{fig:conc_ftle}       
\end{figure*}

Figure \ref{fig:conc_ftle} shows the concentration maps of particles $C$ normalized by the number of particles at each time $t$, for $Re=$1187. During the pulsatile cycle, the concentration in the leading edge of the vortex becomes larger. By comparing figures \ref{fig:conc_ftle}  and \ref{fig:FTLE_barrera1}(c), large concentration values correspond to the $\Lambda^-$ ridge of the FTLE fields.

\subsection{Residence time maps}
\label{sec:rt}

\begin{figure*}[htb]
	\centering
	\resizebox{0.95\textwidth}{!}{%
		\includegraphics{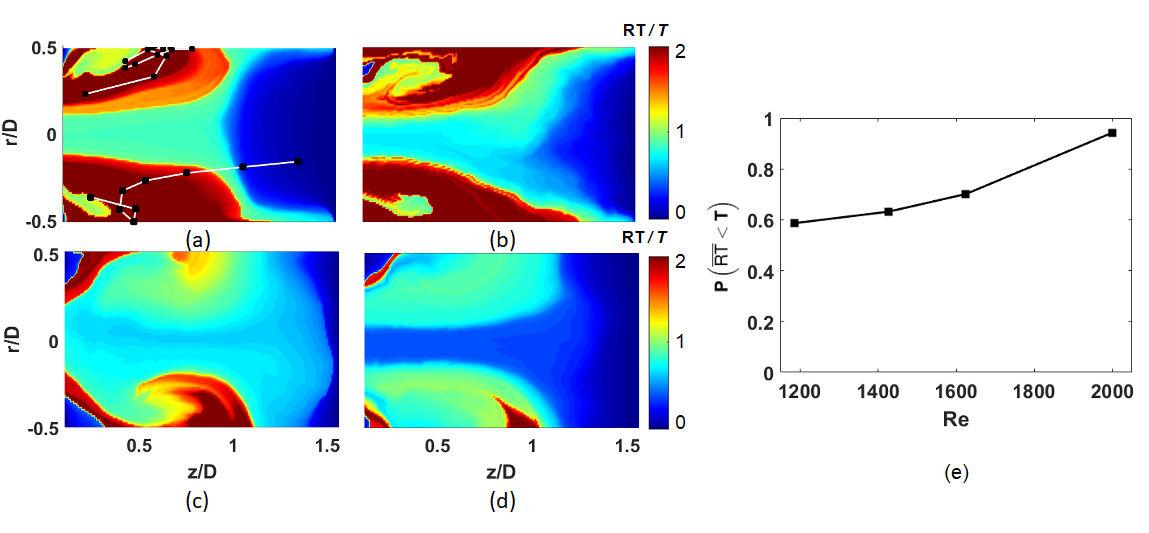}}
	\centering
	\caption{Residence time maps, RT($t_0$=0), for (a) $Re$=1187, (b) $Re$=1427, (c) $Re$=1625 and (d) $Re$=1999. Each pixel represents the fluid particle initial position and the color value represents the time spent by the particle within the ROI normalized by $T$. In (a) particles trajectories corresponding to particles initial positions in the center of the vortex and in the edge of the vortex are shown near the lower and upper wall respectively. (e) Proportion of particles which leave the ROI before a cycle is finished $P(\mathrm{\overline{RT}}<T)$ as function of $Re$.}
	\label{fig:rt_maps}       
\end{figure*}

Figure \ref{fig:rt_maps} shows the RT maps for all the experiments, where  $t_0=0$ is the initial time. The color scale of the RT map corresponds to the residence time normalized by the pulsatile period. Figure \ref{fig:rt_maps} shows that the central jet presents lower residence times compared to the recirculation region for all experiments carried out in this work. For example, for $Re$=1999 (fig. \ref{fig:rt_maps}(d)), RT$<$0.3$T$ in the central jet while in the recirculation region RT$>$0.7$T$. Moreover, figure \ref{fig:rt_maps}(a) shows the trajectory of a particle located in the center of the vortex (lower wall) and in the vortex periphery (upper wall), evidencing two different dynamics within the vortex. Finally, fig. \ref{fig:rt_maps}(e) shows that the proportion of particles that leave the ROI before a period $T$ increases with $Re$. For $Re$ = 1187, 60\% of particles left the region, while 95\% of particles left for Re = 1999.

\subsection{Final position maps}
\label{sec:rf}

\begin{figure*}[htb]
	\centering
	\resizebox{0.95\textwidth}{!}{%
		\includegraphics{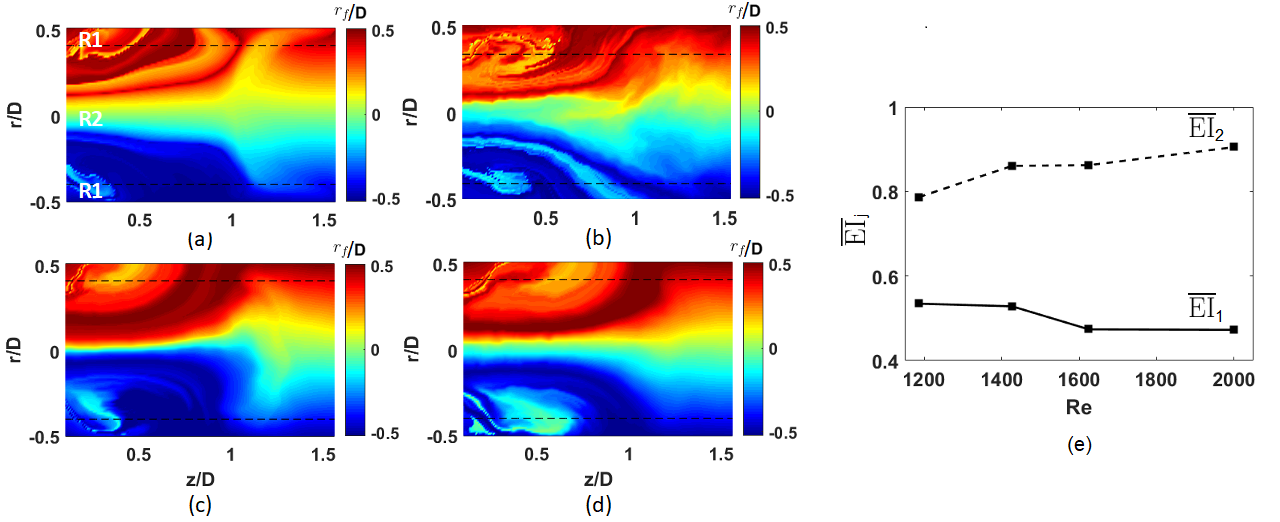}}
	\centering
	\caption{Final position maps, $r_f(t_0=0)$, for (a) $Re$=1187, (b) $Re$=1427, (c) $Re$=1625 and (d) $Re$=1999. Each pixel represents the fluid particle initial position and the color value represents its final position in $r$. Subregions $R1$ and $R2$ are delimited by a dotted line that runs through the center of the vortex. In (e), flow exchange between subregions is measured by $\overline{\mathrm{EI}}_1$ (solid line) and $\overline{\mathrm{EI}}_2$ (dashed line) as a function of $Re$.}
	\label{fig:Rout}       
\end{figure*}

Figure \ref{fig:Rout} shows the $r_f$ maps for the initial time $t_0=$0. Based on the $r_f$ maps for all the  initial conditions, we found that there is no flow exchange between the two halves of the ROI delimited by $r=$0. Therefore, the number of particles in each region does not change substantially. The vortex is the main structure transporting particles in the radial direction. Consequently,  we chose the two subregions $R_1$ and $R_2$ delimited by a line that runs through the center of the vortex and parallel to the wall of the tube, see Figs. \ref{fig:Rout} (a)-(d).

Figures \ref{fig:Rout} (a)-(d), show that most of the flow initially in either the jet or the recirculation region, remains in the same region. However, the mixing occurs within the vortex. For instance, particles initially in the centre of the vortex have final positions close to the jet i.e. subregion $R_2$. While particles initially close to the periphery of the vortex present layers of different $r_f$ values, meaning that its final position is close to the walls. 

Figure \ref{fig:Rout}(e) shows that as $Re$ increases,  $\overline{\mathrm{EI}}_1$ decreases while $\overline{\mathrm{EI}}_2$ increases. 
Therefore, $R_1$ receives more particles from $R_2$, while $R_2$ receives less particles from $R_1$. This constitutes R1 as the main region responsible for increasing the mixing. These parameters can be used as an indicator to quantify the mixing and complete the description made with the RT maps, $r_f$ maps, concentration maps and $\Lambda^-$ fields.

\section{Discussion} \label{sec:discus}

This work studies the mixing of a pulsatile flow in a tube with axisymmetric constriction. To achieve this, we used Lagrangian-based methods to describe mixing processes, the final state of the mixing, and also quantify the mixing. Specifically, we calculated the FTLE fields, concentration maps $C$, and RT maps, and in order to include a spatial description of the particle exchange, we introduced the final position maps $r_f$.

We found that the ridges of  $\Lambda^+$ and $\Lambda^-$  fields organize the flow, as they delimit the vortices and divide the ROI into two distinct regions \cite{shadden_2005, vetel, espa, cagney_2016}. In particular, $\Lambda^+$ ridges, delimit the trailing edge while $\Lambda^-$ ridge delimits the leading edge of the vortex. The flow entering the ROI does not cross the $\Lambda^+$ ridge, separating the flow between successive periods. Similar results were obtained by Shadden et al.\cite{shadden-exp} and Shadden and Taylor \cite{Shadden2008} where LCS predicts the vortex boundaries which organize the flow that enters and leaves the vortex.

By comparing figures \ref{fig:FTLE_barrera1}(c) and \ref{fig:conc_ftle} , we established the connection between the $\Lambda^-$ ridges and the $C$ maps. High concentration values correspond with the $\Lambda^-$ ridge because  it is the attractive LCS. At $t=$0.25$T$ of fig. \ref{fig:conc_ftle}, particles at $z<0.7D$ correspond to particles upstream the leading edge of the vortex (see $\Lambda^-$ ridge in figure \ref{fig:FTLE_barrera1}(c)). As the flow evolves, these particles move towards this edge.  However, particles initially located at $z>$0.7$D$ show different behaviors: particles close to the vortex are attracted by its leading edge, while the others leave the ROI. Finally, the large difference in concentration near the front suggests that an intense mixing occurs locally  \cite{cagney_2016, prat}.

FTLE fields and $C$ maps give an insight into how the mixing occurs, however they do not give the  final state of the mixing or provide quantitative information on mixing. To overcome this issue, we complete this information with RT and $r_f$ maps.


A comparison of the RT maps (fig. \ref{fig:rt_maps}) with the FTLE fields (fig. \ref{fig:FTLE_barrera1}) shows that regions with different dynamics also have different residence times. For instance, the residence times in the centerline are approximately RT=[0.8, 0.8, 0.5, 0.3]$T$ for $Re$=[1187, 1427, 1625, 1999], respectively. We observed that the mean RT in the recirculation region is approximately 2 times larger than the RT in the centerline. Moreover, we observed that this ratio does not change with $Re$. This leads to the accumulation of low velocity flow in the recirculation region and close to the constriction. Jeronimo et al. \cite{jeronimo2019,jeronimo2020} observed analogous qualitative behavior, although RT values are not comparable due to differences between the $Re$ values and the geometries used. 

RT maps also highlighted regions with different dynamics within the vortex. Figure \ref{fig:rt_maps}(a) shows that in the center of the vortex, approximately at $z$=0.25$D$ and $\mid r \mid=$0.4$D$, residence times are lower than in the rest of the vortex. In order to study this difference in detail, we show the trajectories of representative particles initially at the center of the vortex (see fig. \ref{fig:rt_maps}(a) against the lower wall) and from the edge of the vortex (see fig. \ref{fig:rt_maps}(a) against the upper wall). We observed that the particles initially in the center of the vortex move together with the vortex and leave the ROI before the particles initially in the edge, which move towards the walls.

In blood flows, the information given by RT maps can be related to flow stagnation. Previous works \cite{zarins2017pathophysiology} show that flow stagnation is given by high residence time. This can lead to the aggregation of platelets and blood cells, and a low transport of inhibitors, leading to the formation of  thrombosis. 

Previous works describe the final position of particles by lagrangian techniques \cite{dilabbio_2018,hendabadi, jeronimo2019}. Usually final position maps are represented by color-coded subregions of the ROI. However, in this work we used $r_f$ maps which show the exact particle final position in $r$ coordinate. These maps are more appropriate for our purposes since the $r$ coordinate is normal to the boundary between the main structures of the flow: vortex and jet. Then, $r_f$ maps are a reliable representation of the final state of the mixing between these structures.

These maps combined with RT maps, give the spatial and temporal final state of the flow. For instance, we analyze spatio-temporal final state for particles initially close to  $z$=0.25D and $\mid r \mid=$0.4D, by comparing  figs. \ref{fig:rt_maps} and \ref{fig:Rout}. These results show that those particles which remain less time in the ROI, have final positions either within the center of the vortex or the central jet (approximately -0.2D$<r<$0.2D). On the other hand, the particles that were initially at the edge of the vortex show the highest residence time (see fig. \ref{fig:rt_maps}) and its final positions are close to the walls (see fig. \ref{fig:Rout}). We conclude that for all the values of $Re$, the $r_f$ and RT maps delimit the same regions within the ROI. The regions with higher residence time have a  final position close to the walls, whereas the regions with lower residence time have $r_f$ in the central jet.

By calculating the $P(\mathrm{\overline{RT}}<T)$ and $\mathrm{\overline{EI}}$ parameters, we were able to quantify the mixing as $Re$ increases. Figure \ref{fig:rt_maps}(e) showed that residence times decrease as $Re$ increases, which means that mixing is increased with $Re$.  

Although this is an expected result, the incidence of each structure in mixing is not self-evident, considering that the flow is not turbulent. However, further analysis of  $\mathrm{\overline{EI}}$ gives an insight on the incidence of the vortex in the mixing. For instance, fig. \ref{fig:Rout}(e) shows that $\mathrm{\overline{EI}}_1$ decreases and $\mathrm{\overline{EI}}_2$ increases with $Re$. This means that  $R_1$ receives more particles from $R_2$ as $Re$ increases. Since $R_1$ is within the vortex region, these results shows that the vortex is the main structure responsible for the mixing between regions.

Based on previous works \cite{gharib, Barrere2020}, we suggest this mixing behaviour is due to circulation implications on vortex shedding. In a flow being pushed through a constriction, vortex shedding occurs after reaching a specific value of circulation, while any excess of circulation goes to a trail of vorticity. This mechanism  explains that as $Re$ is increased, vortex area is enlarged. Therefore, the increment of $Re$ increases the proportion of particles that attach to the vortex finally leading to an increment of particles in $R_1$. Noteworthy, the introduction of $\mathrm{\overline{EI}}_j$ parameter provides a direct and quantitative measurement of the effects caused by the aforementioned mechanism. 

Overall, we used LCS and Concentration maps to explain the behavior of the main flow structures -i.e., Vortex and central jet – and the mechanism underlying mixing. However, these Lagrangian-based methods do not provide quantitative information on mixing. Since mixing is an emergent spatiotemporal phenomenon, it is difficult to quantify and depends on each application. We consider that in this work, a step was made in giving a comprehensive description and quantification of mixing in our specific system. Specifically, the introduction of $r_f$ maps and its combination with RT maps and $P(\mathrm{\overline{RT}}<T)$ and $\mathrm{\overline{EI}}_j$ parameters provide a set of descriptors to study and measure mixing as a function of Reynolds number.

Future studies should focus on applying these methods in more specific hematological situations. For instance, a possible application of the $P(\mathrm{\overline{RT}}<T)$ parameter is to measure the time of a process of interest. Previous works specify the relevance of thrombi transport times \cite{Rayz_2010}, or the residence time of a flow in a region in pathological conditions \cite{dilabbio_2018, arzani_stagnation}. Thus, RT maps, and particularly $P(\mathrm{\overline{RT}}<T)$, can be used to measure the time of such processes. Similarly, the combination of RT maps and $r_f$ can be useful in applications involving, for example, the transport of blood particles or drug delivery. These problems require identifying specific particles, finding stagnation points, computing particle concentration in different regions, and exchanging particles between specific regions.



\section{Conclusions} \label{sec:conc}

In this work, we used FTLE fields, concentration maps, RT residence time maps, and $r_f$ final position maps as tools to describe the mixing of a pulsatile flow in a tube with an axisymmetric constriction. 
The evolution of how mixing occurs was given by the evolution of  FTLE fields and $C$ maps. The FTLE fields identified a set of LCS that separates the vortex from the central jet and separates the flow between successive periods and sates the vortex leading edge as the structure where particles agglomerate the most. 


We get an insight into the final state of the mixing through RT and $r_f$ maps. These maps are scalar fields computed from particle trajectories and capable of giving information about the topology and mixing of the flow. 
Based on them, we calculated the  $P(\mathrm{\overline{RT}}<T)$ and $\mathrm{\overline{EI}}_j$ parameters to quantify the mixing with  $Re$.  These parameters confirmed that the mixing increases as $Re$ increases and provided an interpretation of how this occurs. For instance, an analysis of $\mathrm{\overline{EI}}_j$, RT, and $r_f$ maps showed that the vortex is the main responsible in mixing process and gives the basis to suggest a mechanism behind this behavior. Finally, we conclude that by cross-referencing the information obtained by each of these tools, we accomplished a better understanding of the mixing processes.

\section{Acknowledgements}
N.B. acknowledges the funds granted by ANII (Agencia Nacional de Investigaci{\'o}n e Innovaci{\'o}n, Uruguay) as part of the Doctoral scholarship POSNAC-2015-1-109843. This research was supported by CSIC (Comisi{\'o}n Sectorial de Investigaci{\'o}n Cient{\'i}fica (CSIC), Uruguay), through the 2016 I+D project ``Estudio din\'amico de un flujo puls\'atil y sus implicaciones hemodin\'amicas vasculares'' and CSIC's group grant ``CSIC2018 - FID13 - grupo ID 722'' and  by PEDECIBA, Uruguay. 

\section{Declarations}

\textbf{Ethical approval.} Not applicable. This article does not contain any studies with human or animal subjects. 

\textbf{Competing interests. }The authors have no conflicts of interest to declare that are relevant to the content of this article.

\textbf{Author's Contribution.} Conceptualization: Nicasio Barrere, Javier Brum, Gustavo Saras\'ua, Cecilia Cabeza; Data Curation: Nicasio Barrere; Formal Analysis: Nicasio Barrere, Javier Brum, Gustavo Saras\'ua, Cecilia Cabeza; Funding acquisition: Javier Brum, Cecilia Cabeza; Investigation: Nicasio Barrere, Javier Brum, Maximiliano Anzibar, Felipe Rinderknecht, Cecilia Cabeza; Methodology: Nicasio Barrere, Javier Brum, Gustavo Saras\'ua, Cecilia Cabeza; Project Administration: Javier Brum, Cecilia Cabeza;  Resources: Javier Brum, Cecilia Cabeza; Software: Nicasio Barrere; Supervision: Nicasio Barrere, Javier Brum, Cecilia Cabeza; Validation: Nicasio Barrere, Javier Brum, Gustavo Saras\'ua, Cecilia Cabeza; Visualization: Nicasio Barrere; Writing - original draft: Nicasio Barrere; Writing - review and editing: Nicasio Barrere, Javier Brum, Gustavo Saras\'ua, Cecilia Cabeza;

\textbf{Funding.} Experimental resources funding provided by CSIC (Comisi{\'o}n Sectorial de Investigaci{\'o}n Cient{\'i}fica (CSIC), Uruguay), through the 2016 I+D project ``Estudio din\'amico de un flujo puls\'atil y sus implicaciones hemodin\'amicas vasculares''. 

\textbf{Availability of data and materials. }
The datasets generated during and/or analysed during the current study are available from the corresponding author on reasonable request.
\appendix
\section{Additional results}
\label{sec:app}

\begin{figure*}[htb]
	\centering
	\resizebox{0.73\textwidth}{!}{%
		\includegraphics{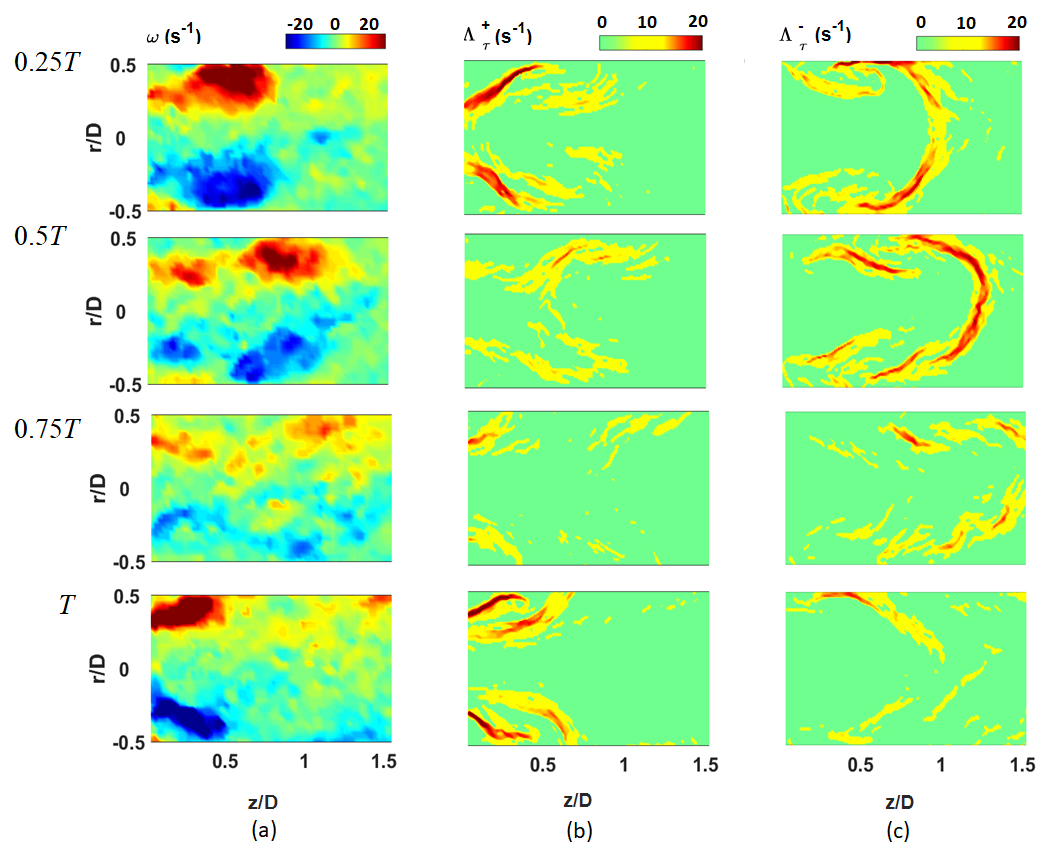}}
	\centering
	\caption{For $Re$=1427, at $t$=[0.25 0.5 0.75 1]$T$ (a) vorticity field $\omega \,(s^{-1})$; (b) positive FTLE fields $\Lambda^+\, (s^{-1})$;  (c) negative FTLE fields $\Lambda^-\, (s^{-1})$. Both FTLE fields were computed with $\tau$=0.25$T$.}
	\label{fig:FTLE_Re1427}       
\end{figure*}

\begin{figure*}[h!]
	\centering
	\resizebox{0.70\textwidth}{!}{%
		\includegraphics{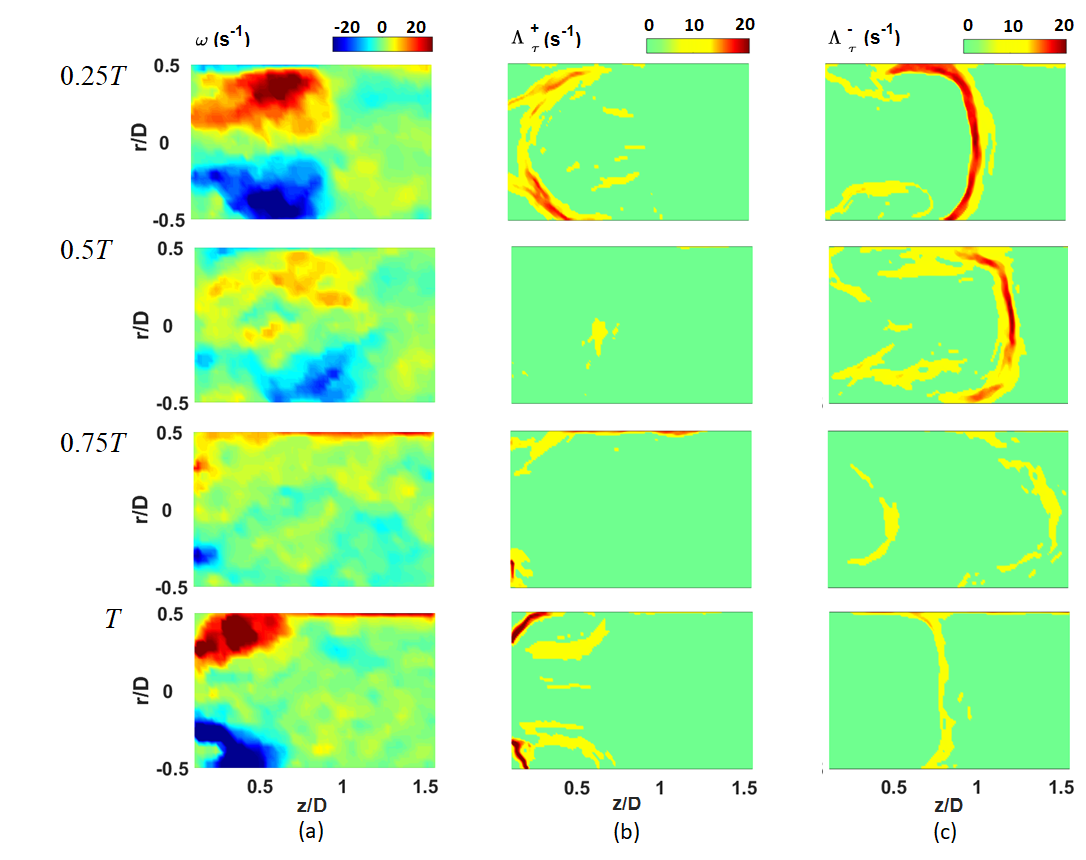}}
	\centering
	\caption{For $Re$=1625, at $t$=[0.25 0.5 0.75 1]$T$ (a) vorticity field $\omega \,(s^{-1})$; (b) positive FTLE fields $\Lambda^+\, (s^{-1})$;  (c) negative FTLE fields $\Lambda^-\, (s^{-1})$. Both FTLE fields were computed with $\tau$=0.25$T$.}
	\label{fig:FTLE_Re1625}       
\end{figure*}

\begin{figure*}[htb]
	\centering
	\resizebox{0.70\textwidth}{!}{%
		\includegraphics{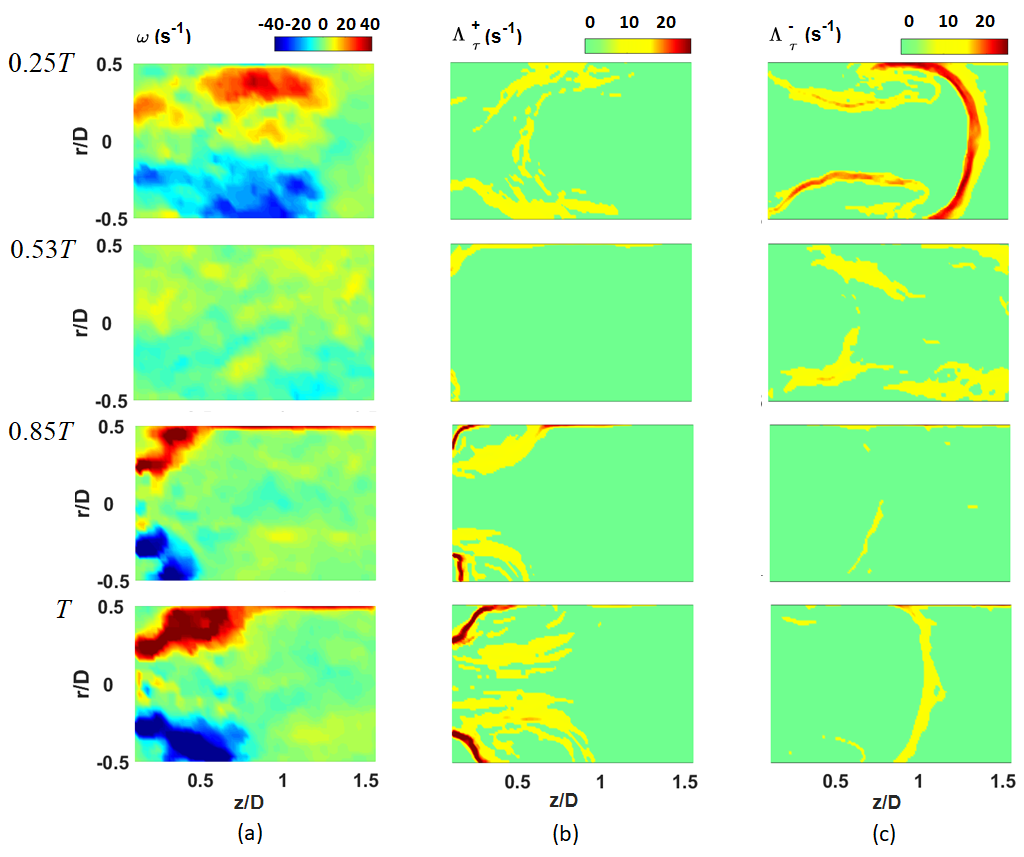}}
	\centering
	\caption{For $Re$=1999, at $t$=[0.25 0.53 0.85 1]$T$ (a) vorticity field $\omega \,(s^{-1})$; (b) positive FTLE fields $\Lambda^+\, (s^{-1})$;  (c) negative FTLE fields $\Lambda^-\, (s^{-1})$. Both FTLE fields were computed with  $\tau$=0.25$T$.}
	\label{fig:FTLE_Re1999}       
\end{figure*}

Figures \ref{fig:FTLE_Re1427}, \ref{fig:FTLE_Re1625} and \ref{fig:FTLE_Re1999} show the vorticity fields, $\Lambda^+$ and $\Lambda^-$  for $Re$=1427, $Re$=1625 and $Re$=1999, respectively. In all cases, we observed structures analogous to those in fig. \ref{fig:FTLE_barrera1}, with $\Lambda^+$ and $\Lambda^-$ ridges which delimit the boundaries of the vortex. As $Re$ increases, the flow becomes disorganized, particularly in the deceleration phase, and the propagation velocity of the flow increases\cite{Barrere2020}. For $Re=$1999 (fig.\ref{fig:FTLE_Re1999}), the phases of the vortex are represented by $t=$0.25$T$, $t=$0.53$T$, $t=$0.85$T$, and $t=T$. At $t$=0.53$T$ we observed disorganized patterns in the vorticity field and the $\Lambda^-$ field due to the vortex tail that had previously left the ROI. Compared with lower $Re$, the ridges of the $\Lambda^-$ field were more filamentous. For $t$=0.85$T$, on the other hand, a new vortex that had not yet detached started to form, hence only the boundary delimited by the ridges of the $\Lambda^+$ field appeared.

\clearpage

\section*{\label{sec:biblio} References}
\bibliography{aipsamp}

\end{document}